\documentclass[epj,final]{svjour}

\usepackage{xcolor}

\usepackage{latexsym}
\usepackage{url}
\usepackage{amsfonts}
\usepackage{amsmath, amssymb}
\RequirePackage{graphicx}
\usepackage{subcaption}

\begin{document}
\title{Tidal disruption effects near black holes and Lambda-gravity}
\author{A.~Stepanian\inst{1}, Sh.~Khlghatyan\inst{1}, V.G. Gurzadyan\inst{1,2}
}                     
%
%
\institute{Center for Cosmology and Astrophysics, Alikhanian National Laboratory and Yerevan State University, Yerevan, Armenia \and
SIA, Sapienza Universita di Roma, Rome, Italy}
\date{Received: date / Revised version: date}
%

\abstract{The tidal disruption of stars in the vicinity of massive black holes is discussed in the context of $\Lambda$-gravity. The latter provides an explanation to the Hubble tension as a possible consequence of two Hubble flows, the local and global ones. The bunch of notions which play role for the considered tidal effect are obtained, along with the rate of the disrupted stars. The role of pulsars is emphasized due to their ability to penetrate up to the horizon of the massive black hole as for them the tidal radius can reach the horizon. Tidal disruption mechanism also can lead to segregation of stars by their mean density vs the distance from the black hole, the denser stars surviving at shorter distances. The interplay of the central gravity field and the repulsive $\Lambda$-term increasing with radius and its certain observational consequences are discussed.}

%
\maketitle

\section{Introduction}

Tidal disruption is a phenomenon which have to occur once a star is getting enough close  to a massive black hole (BH), i.e. when the tidal forces dominate over the star's self-gravity and the star starts to be torn apart. The importance of this inevitable process in the conditions of galactic nuclei was theoretically noticed in \cite{Hills,FR,GO,GO1,GO2,Rees}, and currently the tidal disruption events (TDE) are being actively studied observationally, e.g. \cite{Kom,Saz,Liu,Yao} and references therein. In this paper we analyze certain aspects of the tidal disruption process in the context of $\Lambda$-modified weak-field General Relativity. Various versions of modified gravity are among the approaches to describe the dark sector and the cosmological tensions, e.g.\cite{E,Cap,Char,Dai}.

The equivalency of gravitational field of a sphere and of a point mass located at its center is the principal property of Newtonian gravity. It was proved \cite{G} that, the general function for force $\mathbf{F}(r)$ satisfying this principle has the form 
\begin{equation}\label{1}
\mathbf{F}(r) = \left(-\frac{A}{r^2} + \Lambda r\right)\hat{\mathbf{r}}\ .
\end{equation} 

The important feature of Eq.(\ref{1}) is the fact that in contrast to the first, inverse-squared Newtonian term, the second term, with a cosmological constant $\Lambda$ in McCrea-Milne approach \cite{MM,MM1}, produces a non-force-free field inside the spherical shell. Eq.(\ref{1}) ($\Lambda$-gravity) enables to describe the observational data on galactic halos, groups and clusters of galaxies \cite{G1,GS1,GS2}. It also suggests a solution of the Hubble tension \cite{VTR,R2,R3,Br}, without any additional input parameter, as of a difference between the two Hubble flows, of local and global ones \cite{GS3,GS4}. The local one is defined via Eq.(1) i.e. with a cosmological constant, while the global one is defined via Friedmannian equations .
Other effects in the context of $\Lambda$ gravity are considered in \cite{SK,SKG1,SKG2}, its role in the relative instability of $N$-body gravitating systems has been analyzed in \cite{GKS}. Also, the consideration of $\Lambda$ as of a fundamental physical constant enables one to describe the cosmological evolution within the notion of information \cite{GS3}.

Below we study the influence of the gravity defined by Eq.(1) in the tidal disruption process near the massive black hole situated in a dense stellar system. as typical for the core of galaxies. The main characteristics defining the tidal events and their rates have to be reparameterized respectively. The role of compact stars with higher density such as pulsars regarding the disruption events has to be notable as for them the Roche (tidal) lobe can become smaller than the supermassive black hole's horizon. Then, the tidal disruption effects have to lead to overdensity of dense stars at the very vicinity of the black hole.


\section{Tidal disruption of stars}

The distance from the central massive BH of mass $M_{BH}$ determining the crucial role of the tidal effects for the stability of stars is defined by the so-called Roche limit or tidal radius which is defined as 
\begin{equation*}
r_t = \left(2\frac{M_{BH}}{m_{\ast}}\right)^{1/3}r_{\ast},
\end{equation*}
where  $m_{\ast}$ is the mean stellar mass. For polytropic equations of state $P=K\rho^{(n+1)/n}$ a coefficient $\eta$ depending  on the polytropic index does appear $n$ \cite{MT,M}
\begin{equation}\label{tidalradius}
r_t = \eta^{2/3}\left(\frac{M_{BH}}{m_{\ast}}\right)^{1/3}r_{\ast}.
\end{equation}
Thus, a star at a distance equal or less than determined by Eq.(\ref{tidalradius}) will be destroyed by tidal forces of the massive black hole.

Consider the angular momentum of a star as follows
\begin{equation}
J_{lc}= \sqrt{2GM_{BH}r_{lc}},
\end{equation}
where $\varepsilon$ is the star's orbital binding energy per unit mass and $r_{lc}$ so-called loss-cone radius which itself is defined as the maximum value among the tidal and capture radii $r_{lc}=\text{max}\{r_{t},r_c=8GM_{BH}/c^2\}$. Accordingly, instead of definition of tidal radius one can use the notion of $J_{lc}$ as \cite{MT,M}
\begin{equation}\label{JLC}
r < r_t \equiv J<J_{lc}.
\end{equation}
Similarly, $\theta_{lc}$ is the angle when the orbits with within $\theta < \theta_{lc}$ will pass within $r_{t}$ and is defined as \cite{M}
\begin{align*}
\theta_{lc} &= \sqrt{\frac{r_{lc}}{r}} \quad r<r_h, \\ 
\theta_{lc} &= \sqrt{\frac{r_{lc}r_h}{r^2}} \quad r>r_h,  \\ 
\end{align*}
where $r_h=GM_{BH}/\sigma^2$ is the radius of the sphere of influence of the BH, $\sigma$ is the stellar velocity dispersion.

We consider a spherical system with the star distribution function $f(\pmb{x},\pmb{v})$. Turning from $(\pmb{x},\pmb{v})$ phase space to energy-momentum space $(\varepsilon,J^2)$, the probability of finding a star in a space volume $d\varepsilon dJ^2$ will be
\begin{equation}
N(\varepsilon,J^2)=\int f(\varepsilon, J^2)d^3{x}d^3{v}=4\pi^2 f(\varepsilon,J^2)P(\varepsilon,J^2)d\varepsilon dJ^2,
\end{equation}
where $P$ is the period of the motion. The number of stars $N_{lc}(\varepsilon)$ in the (full) loss cone for a spherical system with isotropic distribution function is given by
\begin{equation}\label{LC}
N_{lc}(\varepsilon)d\varepsilon = 4\pi^2 f(\varepsilon)P(\varepsilon)J_{lc}^2d\varepsilon.
\end{equation}
For simplicity we consider systems of steep power-law core density profiles, i.e. $\rho\sim r^{-\alpha}$ where $\alpha= 2$ corresponds to the isothermal profile. Accordingly, since for simplicity it is assumed that the stellar system (galactic core) includes stars of approximately the same mass $m_{\ast}$, the mass density is related to the number density as $\rho(r)=m_{\ast}\mu(r) $. Thus,  the stellar orbits are isotropic in angular momentum space and therefore can be described by a distribution function which depends only on energy \cite{MT}. Then for the number density of stars $\nu = \nu_0(r_h/r)^{\alpha}$,  the distribution function will be
\begin{equation}\label{DF}
f(\varepsilon) = (2\pi\sigma_h^2)^{-3/2}\nu_0\frac{\Gamma(\alpha + 1)}{\Gamma\left(\alpha-\frac{1}{2}\right)}\left(\frac{\varepsilon}{\sigma_h^2}\right)^{\alpha-3/2}, \quad \frac{1}{2}<\alpha<3.
\end{equation}
Thus, for finite orbits, when the loss cone is full, the stars will be destroyed during one dynamical time which can be regarded as the orbital period, and the flux of stars (rate) is determined by the following expression
\begin{equation}
F^{full}(\varepsilon) = \frac{N_{lc}(\varepsilon)}{P(\varepsilon)}.
\end{equation}

In the next section we will obtain this flux and other parameters in the context of $\Lambda$-gravity, Eq.(1), to reveal the contribution of $\Lambda$.

\section{Modified gravity}

The cosmological constant $\Lambda$ naturally enters into the equation of the weak field as an second term in Eq.(\ref{1}). Therefore, the characteristic radii and dynamical quantities must be modified correspondingly.

First, the orbital binding energy is modified as follows
\begin{equation}\label{bind}
\varepsilon = \frac{GM_{BH}}{2r} + \frac{\Lambda c^2 r^2}{3}.
\end{equation}
Accordingly, the tidal radius will be written
\begin{equation}\label{tid}
r_{t,\Lambda} = \eta^{2/3}\left(\frac{2GM_{BH}}{m_{\ast}G-\Lambda c^2 r_{\ast}^3}\right)^{1/3}r_{\ast}.
\end{equation}
Table \ref{tab1} shows the comparison among the first and second terms of the denominator in Eq.(\ref{tid}) in terms of the star density.

\begin{table}[h]
\caption{$\Lambda$-modified tidal radius for a sample of different type of stars when the central BH is SgrA${}^*$, $M_{BH}= 10^6 M_{\odot}$ \cite{Ant,Tay}.}\label{tab1}
\centering
\begin{tabular}{ |p{2.4cm}||p{2.4cm}||p{2.4cm}||p{2.7cm}| }
\hline
Star& Density [kg][m]$^{-3}$& Compare  $\left(\frac{3\Lambda c^2}{G}/4\pi \rho_{\ast}\right)$& $r_{t,\Lambda}$ [m] \\
\hline
PSR J0348+0432& 4.293542$\times 10^{17}$ & 8.327896 $\times 10^{-44}$& 1.300959  $\times 10^{6}$   \\
\hline
Sun & 1403.817238& 2.547068 $\times 10^{-29}$& 8.765271  $\times 10^{10}$  \\
\hline
Sirius & 578.175614& 6.184310 $\times 10^{-29}$& 1.178100  $\times 10^{11}$ \\
\hline
Procyon	& 225.8598 & 1.583114 $\times 10^{-28}$& 1.611595  $\times 10^{11}$   \\
\hline
Pismis 24-17 & 22.287349& 1.604326 $\times 10^{-27}$& 3.487515  $\times 10^{11}$    \\
\hline
R136a1 & 5.010613& 7.136088 $\times 10^{-27}$& 5.735508  $\times 10^{11}$ 	 \\
\hline
Aldebaran& 1.787027$\times 10^{-2}$ & 2.000875 $\times 10^{-24}$& 3.753979  $\times 10^{12}$    \\
\hline

\end{tabular}
\end{table}

\begin{figure}
\centering
\begin{subfigure}{.5\textwidth}
\centering
\includegraphics[width=.9\linewidth]{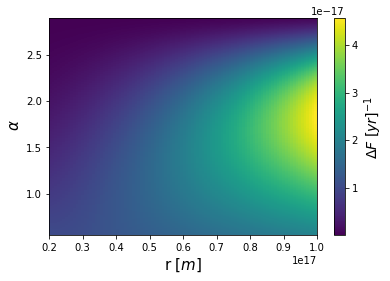}
\caption{}
\label{fig:sub1}
\end{subfigure}%
\begin{subfigure}{.5\textwidth}
\centering
\includegraphics[width=1.\linewidth]{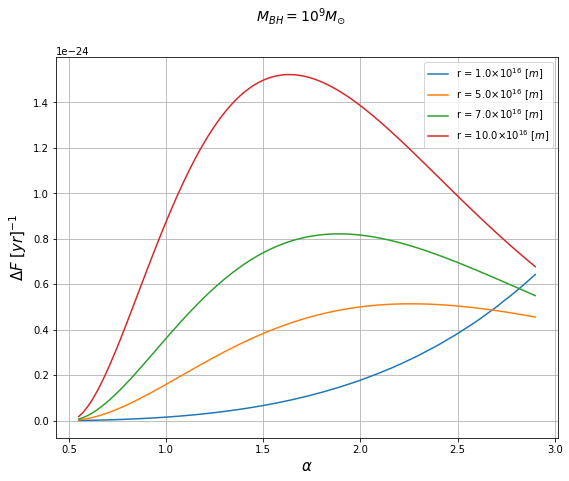}
\caption{}
\end{subfigure}
\caption{(a) The behavior of the disruption flux $\Delta f$ with respect to $r$ and the star number density parameter $\alpha$ for a BH of $10^{9}$ solar masses; (b) The dependence of $\Delta f$ on $\alpha$ for different values of $r$. }\label{DFplot}
\end{figure}

\begin{figure}
\centering
\includegraphics[width=0.6\textwidth]{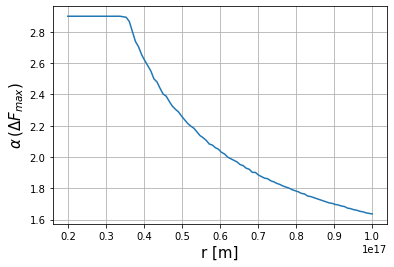}
\caption{The behavior of $\alpha$ versus $r$, when $\Delta f$ is maximal.}\label{DFmax}
\end{figure}

\noindent
Then, in the context of $\Lambda$-gravity the loss-cone angular momentum and orbital period are written as 
\begin{equation}\label{Jlc}
J_{lc,\Lambda} = 2GM_{BH}r_{t,\Lambda}+ \frac{\Lambda c^2}{3}r_{t,\Lambda}^4, \quad P_{\Lambda} = \frac{2 \pi}{\sqrt{\frac{GM_{BH}}{r^3} - \frac{\Lambda c^2}{3}}}
\end{equation}
Using the above modified expressions and Eq.(\ref{DF}) the flux is obtained
\begin{equation}\label{Flux}
f^{full}_{\Lambda}(r)= (2\pi)^{1/2}\frac{\Gamma(\alpha+1)}{\Gamma(\alpha -1/2)}\times\nu_0\times\left(\frac{GM_{BH}}{2r} + \frac{\Lambda c^2 r^2}{3}\right)^{\alpha-3/2}\times\left(\frac{1}{\sigma_h^2}\right)^{\alpha}\times\left[2GM_{BH}r_{t,\Lambda}+\frac{\Lambda c^2}{3}r_{t,\Lambda}^4\right]
\end{equation}
For typical parameters of BH \cite{M}, i.e.  $M_{BH}=10^6M_{\odot}$ and $\sigma = 100\, km\, s^{-1} ,\,\nu_0 = 10^{5}\, pc^{-3},\,\alpha = 1.5$  we obtain the following values:

\begin{align}
f^{full}_{\Lambda}(10^{14})&= 1.671920\times 10^{-3}\, yr^{-1}\\
f^{full}_{\Lambda=0}(10^{14})&= 1.671918\times 10^{-3}\, yr^{-1}.
\end{align} 
It is worth mentioning that, according to Eq.(\ref{Flux}) it turns out that the flux depends strongly on the value of the $\alpha$. The behavior of the flux difference $\Delta f = f_{\Lambda}^{full}-f_{\Lambda=0}^{full}$ with respect to  $\alpha$ and $r$ is shown in Fig.(\ref{DFplot}) for a BH mass $10^9 M_{\odot}$. We also obtain the plot of $\alpha$ versus $r$ for the cases, where $\Delta F$ is the maximal, Fig.(\ref{DFmax}). The plot reveals that for radii smaller than $3.61\times 10^{16}\, m$ the rate of change for $\alpha$ remains almost uniform, while for $r > 3.61\times 10^{16}\, m$ it starts to decrease.

Since, in this case $\Lambda$ is included both in the denominator and the numerator of the flux expression Eq.(\ref{Flux}),  to reveal the contribution of $\Lambda$, one can expand this expression considering the fact that the numerical value of $\Lambda$ is small. In first approximation by $\Lambda$ it has a form 
\begin{equation}\label{ser}
\begin{split}
f_{\Lambda}&=2^{1/6}\pi^{-3/2}\nu_0 \frac{\Gamma(\alpha + 1)}{\Gamma(\alpha-1/2)}\eta^{3/2}r_{\ast}\frac{c^2}{3}\left(\frac{M_{BH}}{m_{\ast}}\right)^{1/3}\left(\frac{1}{r_h}\right)^{-\alpha}(GM_{BH})^{-3/2}\Bigg[2^{13/6-\alpha}\left(\frac{1}{r}\right)^{\alpha-5/2}+\\
&+\Lambda\Big(2^{7/6-\alpha}r_{\ast}^3\frac{c^2}{3}\left(\frac{1}{r}\right)^{\alpha-5/2}\frac{1}{m_{\ast}G}+\eta^{3/2}r_{\ast}\alpha\left(\frac{M_{BH}}{m_{\ast}}\right)^{1/3}\left(\frac{1}{r_h}\right)^{-3}(GM_{BH})^{-\alpha+3/2}+\\
&+\eta^{9/2}2^{14/3-\alpha}r_{\ast}\frac{c^2}{3}\left(\frac{1}{r}\right)^{2\alpha-5}(GM_{BH})^{2\alpha-6}\Big)\Bigg].
\end{split}
\end{equation}

It turns out that different parameters of the system appear in different powers in this expression. Thus, it is important to find out to what extent $\Lambda$ is affecting the expression for the flux. 


As it is clear, the value of $f^{full}$ becomes larger for $\Lambda$-gravity than for Newtonian case, which means that the contribution of $\Lambda$ leads to more stars to be torn apart. The presence of repulsive $\Lambda$-term affects the nature of dynamics of the orbiting stars, especially at larger radii, the notions and concepts of the tidal radius and tidal force.  

\section{Pulsars}

From Table \ref{tab1} one can observe that for pulsars the tidal radius is less than the event horizon of the considered massive BHs. 
Here it is worth mentioning that, by considering $\Lambda$ in the equations of General Relativity, the notion of BH’s event horizon changes drastically. Indeed, for the neutral, static, spherically symmetric case it is obtained by solving the following equation (Schwarzschild-de Sitter metric) \cite{Rind}
\begin{equation}
1- \frac{GM_{BH}}{c^2 r} - \frac{\Lambda r^2}{3} = 0,
\end{equation} 
with the solutions 
\begin{equation}
r_1  = \frac{2}{\sqrt \Lambda} cos\left(\frac{1}{3} cos^{-1}\left(\frac{3GM_{BH} \sqrt \Lambda}{c^2}\right)+\frac{\pi}{3}\right),
\end{equation}
\begin{equation}
r_2 = \frac{2}{\sqrt \Lambda} cos\left(\frac{1}{3} cos^{-1}\left(\frac{3GM_{BH} \sqrt \Lambda}{c^2}\right)-\frac{\pi}{3}\right),
\end{equation}
\begin{equation}
r_3 = -(r_1 + r_2).
\end{equation}
For our case we need to focus only on $r_1$ which is slightly larger than Schwarzschild radius $r_s = 2GM_{BH}/c^2$. 
\begin{equation}
r_1 \approx	\frac{2}{\sqrt \Lambda}\left(\frac{GM_{BH}\sqrt \Lambda}{c^2} + 0 + \left(\frac{GM_{BH}}{c^2}\right)^{\frac{3}{2}}\frac{\Lambda^{\frac{3}{2}}}{6}\right)
\end{equation}
Thus one can state that the pulsar will not undergo tidal disruption before crossing the horizon of galactic BHs.  This fact is notable since pulsars can serve as precise clocks, and not being destroyed by tidal forces, they can act as tracers for time dilation up to the scales of the BH horizon. In the presence of $\Lambda$ the time delay is
\begin{equation}\label{Tdelay}
t_0 = t_f \sqrt{1 - \frac{2GM}{c^2 r} - \frac{\Lambda r^2}{3}},
\end{equation}
where $t_0$  is the proper time shift between two events while the $t_f$ is that of the coordinate time.

\section{Conclusions}

We analyzed the tidal disruption process in the context of $\Lambda$-gravity, Eq.(1). This approach has enabled the description of the galactic halos, dynamics of groups and clusters of galaxies within a unified picture without any free parameter. Also, considering a local Hubble flow defined by Eq.(1), and a global flow defined by Friedmannian equations, one can quantitatively explain the H-tension and obtain strict lower and upper bounds for the
local Hubble parameter. This is also related to the formation mechanisms of the cosmic web and voids \cite{spot}. The second term with the cosmological constant in Eq.(1) is shown to make more unstable (chaotic) the N-body gravitating systems as compared to the case of Newtonian gravity. The study of the tidal disruption mechanism, considered as responsible to the energy release events detected in galactic nuclei within $\Lambda$-gravity thus is also reasonable.  
   
The notions determining the tidal disruptions, the star loss rate which determines the matter accretion and hence the energy release features, is obtained in loss-cone approximation. It is shown that the repulsive $\Lambda$-term tends to increase the tidal disruption rate. 

For pulsars, due to their high matter density, the Roche (tidal) radius is essentially smaller than for main sequence stars and can reach the radius of the horizon of galactic black BHs. Then, pulsars can cross the supermassive BH's horizon without being tidally destroyed, thus outlining their role of precise clocks to trace the time dilation and the space-time metric, e.g. \cite{P1,P2}. Also, the tidal disruption mechanism has to lead to segregation of stars by their mean density vs the distance from the central BH in galactic cores, i.e. as a tendency for storing of higher density stars at very central regions which are "forbidden" for main sequence stars. 

\section{Data Availability Statement} 
Data sharing not applicable to this article as no datasets were generated or analysed during the current study.


\begin{thebibliography}{99}
\bibitem{Hills} J.G. Hills, Nature 254, 295 (1975)
\bibitem{FR} J. Frank, M.J. Rees, MNRAS, 176, 633 (1976)
\bibitem{GO} V.G Gurzadyan, L.M. Ozernoy,  Nature, 280, 214 (1979)
\bibitem{GO1} V.G Gurzadyan, L.M. Ozernoy, A\&A, 86, 315 (1980)
\bibitem{GO2} V.G Gurzadyan, L.M. Ozernoy, A\&A, 95, 39 (1981)
\bibitem{Rees} M.J. Rees, Nature, 333, 523 (1988)
\bibitem{Kom} S. Komossa, Journ. High Energy Astrophys., 7, 148 (2015)
\bibitem{Saz} S. Sazonov, et al, MNRAS, 508, 3820 (2021)
\bibitem{Liu} C. Liu, et al,  arXiv:2206.13494
\bibitem{Yao} Y. Yao, et al, arXiv:2206.12713 
\bibitem{E} M. Eingorn,  ApJ, 825, 84 (2016)
\bibitem{Cap}  S. Capozziello, et al,  MNRAS, 474, 2430 (2018)
\bibitem{Char} G. Chardin, et al,  A\&A, 652, A91 (2021) 
\bibitem{Dai} M.G. Dainotti, et al, ApJ, 912, 150 (2021)
\bibitem{G} V.G. Gurzadyan,  Observatory, 105, 42 (1985)
\bibitem{MM} W.H. McCrea, E.A. Milne, Q. J. Math. 5, 73 (1934)
\bibitem{MM1} E.A. Milne, Q. J. Math.  5, 64 (1934)
\bibitem{G1} V.G. Gurzadyan, Eur. Phys. J. Plus, 134, 14 (2019)
\bibitem{GS1} V.G. Gurzadyan, A. Stepanian,  Eur. Phys. J. C, 78, 632 (2018)
\bibitem{GS2} V.G. Gurzadyan, A. Stepanian,  Eur. Phys. J. C 79, 169 (2019)
\bibitem{VTR}  L. Verde,  T. Treu,  A.G. Riess, Nature Astronomy, 3, 891 (2019)
\bibitem{R2}  A.G. Riess, Nature Review Physics, 2, 10 (2020)
\bibitem{R3}  A.G. Riess, et al., arXiv:2112.04510 (2021)
\bibitem{Br} D. Brout, et al,	arXiv:2202.04077 (2022)
\bibitem{GS3} V.G. Gurzadyan, A. Stepanyan, Eur. Phys. J. Plus,   136, 235 (2021)
\bibitem{GS4}  V.G. Gurzadyan, A. Stepanian, A\&A, 653, A145 (2021)
\bibitem{SK} A. Stepanian, Sh. Khlghatyan, Eur. Phys. J. Plus 135, 712 (2020)
\bibitem{SKG1} A. Stepanian, Sh. Khlghatyan, V.G. Gurzadyan, Eur. Phys. J. C, 80, 1011 (2020)
\bibitem{SKG2} A. Stepanian, Sh. Khlghatyan, V.G. Gurzadyan, Eur. Phys. J. Plus, 136, 127 (2021)
\bibitem{GKS} V.G. Gurzadyan, A.A. Kocharyan, A. Stepanian, Eur. Phys. J. C 80, 24 (2020)


\bibitem{MT} J. Magorrian, S. Tremaine, MNRAS, 309.2, 447 (1999) 
\bibitem{M} D. Merritt, Class. Quant. Gravity, 30, 244005 (2013)

\bibitem{Ant} J. Antoniadis et al, Science, 340, 448 (2013)
\bibitem{Tay} R.J. Tayler, {\it The Stars: Their Structure and Evolution}, (Cambridge University Press, 1994) 

\bibitem{Rind}W. Rindler, {\it Relativity: Special, General and Cosmological}, (Oxford University Press, 2006)


\bibitem{spot} V.G. Gurzadyan et al,  A\&A, 566, A135 (2014)

\bibitem{P1} K. Liu, ApJ, 747, L1 (2012)
\bibitem{P2} D. Psaltis, W. Norbert and M. Kramer, ApJ, 818, 121 (2016)
 


\end{thebibliography}
\end{document}